# The influence of cluster emission and the symmetry energy on neutron-proton spectral double ratios

Yingxun Zhang[1,2,3], P. Danielewicz[1,2,4], M. Famiano[5], Zhuxia Li[3], W.G. Lynch[1,2,4], M.B. Tsang[1,2,4*]

[1] Joint Institute of Nuclear Astrophysics, Michigan State University, East Lansing, MI 48824, USA,
[2] National Superconducting Cyclotron Laboratory, Michigan State University, East Lansing, MI 48824, USA,
[3] China Institute of Atomic Energy, P.O. Box 275 (18), Beijing 102413, P.R. China,
[4] Physics and Astronomy Department, Michigan State University, East Lansing, MI 48824, USA.
[5] Physics Department, Western Michigan University, Kalamazoo, MI, USA.

Abstract

The emissions of neutrons, protons and bound clusters from central $^{124}$Sn+$^{124}$Sn and $^{112}$Sn+$^{112}$Sn collisions are simulated using the Improved Quantum Molecular Dynamics model for two different density-dependent symmetry-energy functions. The calculated neutron-proton spectral double ratios for these two systems are sensitive to the density dependence of the symmetry energy, consistent with previous work. Cluster emission increases the double ratios in the low energy region relative to values calculated in a coalescence-invariant approach. To circumvent uncertainties in cluster production and secondary decays, it is important to have more accurate measurements of the neutron-proton ratios at higher energies in the center of mass system, where the influence of such effects is reduced.

- Corresponding author: tsang@nscl.msu.edu

Information about the Equation of State (EOS) of asymmetric nuclear matter can improve our understanding of the radii and moments of inertia, maximum masses [1-3], crustal vibration frequencies [4], and cooling rates [3,5] of neutron stars, which are currently being investigated at ground-based and satellite observatories. Recent X-ray observations have been interpreted as requiring an unusually repulsive equation of state for neutron matter [6]. It is important to determine whether such interpretations are supported by laboratory measurements. Measurements of isoscalar collective vibrations, collective flow and kaon production in energetic nucleus-nucleus collisions have constrained the equation of state for symmetric matter for densities ranging from normal saturation density to five times saturation density [7-9]. On the other hand, the extrapolation of the EOS to neutron–rich matter depends on the density dependence of the nuclear symmetry energy, for which there are comparatively few experimental constraints [10].

Various probes in reaction experiments have been found to be sensitive to the symmetry energy term of the equation of state. These include isoscaling [11-13], isospin diffusion [14], neutron to proton (n-p) ratios ($R_{n/p}$) [15-17], neutron and proton flow [18], $\pi^+/\pi^-$ ratios, and $\pi^+$ and $\pi^-$ flow [19, 20]. In this paper, we focus on the ratio $R_{n/p}$ of pre-equilibrium neutron over proton spectra. The ratio $R_{n/p}$ is enhanced by the repulsion of neutrons and attraction of protons produced by the symmetry mean field potential, which changes over time with the evolving density and asymmetry of the system [16,17,21].

Experimentally, neutrons and protons are usually measured utilizing two different detection systems with different energy calibrations and efficiencies. An accurate determination of absolute detection efficiencies for neutrons is rather difficult. For these reasons, the first comparison [15] of neutron to proton spectra used the double ratio,

$$DR(n/p) = R_{n/p}(A)/ R_{n/p}(B) = \frac{dM_n(A)/dE_{c.m.}}{dM_p(A)/dE_{c.m.}} \cdot \frac{dM_p(B)/dE_{c.m.}}{dM_n(B)/dE_{c.m.}},$$

constructed by measuring the energy spectra, $dM/dE_{C.M.}$, of neutrons and protons for two systems A and B characterized by different isospin asymmetries.

The sensitivity of $R_{n/p}$ to the symmetry energy has been studied in the past decade using the Boltzmann Uhling Uhlenbeck equation [16, 17, 21], which does not predict cluster formation. Conservation laws dictate that the inclusion of nucleons from $\alpha$ particles and other relatively symmetric clusters can significantly modify the values for $R_{n/p}$ [22]. Thus, it is important to examine the effect of clusters on n-p ratios constructed in dynamical models. To understand this issue, we have performed simulations with the Improved Quantum Molecular Dynamics transport model [23, 24] using two equations of state that differ in their symmetry energy terms.

Within the ImQMD model, nucleons are represented by Gaussian wavepackets. The mean fields acting on these wavepackets are derived from an energy functional with the potential energy U that includes the full Skyrme potential energy with just the spin-orbit term omitted:

$$. \qquad (1)$$

In the above, is the Coulomb energy. The nuclear contributions are represented in a local form with

$$U_{\rho,md} = \int u_{\rho,md} d^3r \qquad (2)$$

and the density potential of symmetry energy $\frac{C_s}{2}(\frac{\rho}{\rho_0})^{\gamma_i}\delta^2\rho$ is included in

$$u_\rho = \frac{\alpha}{2}\frac{\rho^2}{\rho_0} + \frac{\beta}{\eta+1}\frac{\rho^{\eta+1}}{\rho_0^\eta} + \frac{g_{sur}}{2\rho_0}(\nabla\rho)^2 + \frac{g_{sur,iso}}{\rho_0}[\nabla(\rho_n - \rho_p)]^2 + \frac{C_s}{2}(\frac{\rho}{\rho_0})^{\gamma_i}\delta^2\rho + g_{\rho\tau}\frac{\rho^{8/3}}{\rho_0^{5/3}}. \qquad (3)$$

where, the asymmetry is defined as $\delta = (\rho_n - \rho_p)/(\rho_n + \rho_p)$, and $\rho_n$ and $\rho_p$ are the neutron and proton densities, respectively. The energy density associated with the mean-field momentum dependence is given by

$$u_{md} = \frac{1}{2\rho_0}\sum_{N_1,N_2=n,p}\frac{1}{16\pi^6}\int d^3p_1 d^3p_2 f_{N_1}(\vec{p}_1)f_{N_2}(\vec{p}_2) 1.57[\ln(1 + 5\times 10^{-4}(\Delta p)^2)]^2, \qquad (4)$$

where $f_N$ are nucleon Wigner functions, $\Delta p = |\vec{p}_1 - \vec{p}_2|$. The energy is in MeV and momenta are in MeV/c. The associated mean fields acting on the wavepackets can be found in Ref. [25]. In this work, the values of α=-356 MeV, β=303 MeV and η=7/6 are employed, corresponding to an isoscaler compressibility constant of K=200 MeV. Other parameter values are $g_{sur}$=19.47 MeVfm$^2$, $g_{suriso}$=-11.35 MeVfm$^2$, $C_s$=35.19 MeV, and $g_{\rho\tau}$=0 MeV. These calculations use isospin-dependent in-medium nucleon-nucleon scattering cross sections in the collision term and employ Pauli blocking effects that are described in [23,24, 26]. Cluster yields are calculated by means of the coalescence model, widely used in QMD calculations, in which particles with relative momenta smaller than $P_0$ and relative distances smaller than $R_0$ are coalesced into one cluster. In the present work, the values of $R_0 = 3.5 \, fm$ and $P_0 = 250 \, MeV/c$ are employed.

As a consequence of the above assumptions, the symmetry energy per nucleon employed in the simulations is a sum of kinetic and interaction terms:

$$E_{sym}(\rho)/A = \frac{1}{3}\frac{\hbar^2}{2m}\rho_0^{2/3}\left(\frac{3\pi^2}{2}\frac{\rho}{\rho_0}\right)^{2/3} + \frac{C_s}{2}\left(\frac{\rho}{\rho_0}\right)^{\gamma_i}, \qquad (5)$$

where m is the nucleon mass. The symmetry energy for the present ImQMD calculations (solid lines) is plotted as a function of density in Figure 1 for $\gamma_i$=0.5 and 2. The symmetry energy values increase with decreasing $\gamma_i$ at subsaturation densities while the opposite is true at suprasaturation densities. At any density, higher values of symmetry energy tend to drive systems more rapidly towards isospin symmetry, resulting in higher values of $R_{n/p}$ for neutron rich systems.

We have performed calculations of central collisions at an impact parameter of b=2 fm and an incident energy of 50 MeV per nucleon for two systems: A=$^{124}$Sn+$^{124}$Sn and B=$^{112}$Sn+$^{112}$Sn. In central collisions at this incident energy, particles are mostly emitted when the system expands and breaks up at sub-saturation densities. While the ratios of total emitted neutron over total emitted proton numbers are fixed by conservation laws, the important symmetry energy information is contained in the pre-equilibrium emission of nucleons from the early asymmetric system, which dominates at high center of mass (C.M.) energies at $\theta_{C.M.} \approx 90^0$. The right and left panels of Figure 2 show $R_{n/p}$(124) and $R_{n/p}$(112) for the $^{124}$Sn+$^{124}$Sn and $^{112}$Sn+$^{112}$Sn collisions, respectively, as a function of the C.M. energy of nucleons emitted at 70°≤$\theta_{C.M.}$≤110°. The open and solid symbols represent $R_{n/p}$ values calculated using the softer ($\gamma_i$=0.5) and stiffer ($\gamma_i$=2) density-dependent symmetry terms, respectively. As expected, more pre-equilibrium neutrons get emitted from the neutron rich $^{124}$Sn+$^{124}$Sn system. Relatively more pre-equilibrium neutrons are also emitted in the calculations with the softer density-dependent symmetry energy because the emission occurs predominantly at sub-saturation density. The uncertainties for these calculations in Figure 2 and in the subsequent figures are statistical.

To facilitate comparisons to existing and future transport model calculations, we restrict our calculations to b=2 fm in this letter. (Calculations with experimental multiplicity gates imposed on impact parameter averaged events yield results which are consistent with those for b=2 fm, within statistical uncertainties.) While the various uncertainties of the calculations are too large to allow for rigorous comparisons with data, (see the discussion below), some comparisons with data [15] will be shown to provide context for the discussion. The shaded regions in the left panel of Figure 3 represent the range, determined by uncertainties in the simulations, of predicted double ratios $DR(n/p)=R_{n/p}(124)/R_{n/p}(112)$, as a function of the nucleon center of mass energy, for two different density dependencies of the symmetry potential: for $\gamma_i=0.5$ (upper shaded region) and for $\gamma_i=2$ (lower shaded region). All calculated results exceed the no-sensitivity limit of $DR(n/p)=N_A Z_B/(N_B Z_A)=1.2$ (dotted lines) given by conservation laws. As expected, the double ratios $DR(n/p)$ are higher for $\gamma_i=0.5$, which yields weaker dependence of symmetry energy on density. The measured [15] double ratios $DR(n/p)$ are plotted as solid stars for comparison. Both calculations yield results, which increase in values with kinetic energy as observed in the data.

To examine the influence of sequential decays, we have simulated decays of fragments created in the collisions using the Gemini code [27]. Sequential decays mainly enhance the single ratios for low energy protons and neutrons, but such effects are largely suppressed in the double ratios. This underscores the utility of double ratios for comparisons of calculated and measured neutron and proton spectra at energies where the secondary decay contributions may be small but uncertain.

For models that do not include clusters such as the BUU calculations discussed below, coalescence-invariant $DR(n/p)$ are used. These double ratios are constructed by including all neutrons and protons emitted at a given velocity, regardless of whether they are emitted free or within a cluster. The data, shown as solid stars in the right panel of Figure 3, increase monotonically from the no-sensitivity limit $DR(n/p) \approx 1.2$ and attain values at large $E_{C.M.}$ consistent with those shown in the left panel of Figure 3 for free nucleons. The corresponding coalescence-invariant n-p double ratios using the fragments produced in the ImQMD simulations are plotted as shaded regions in the right panel in Figure 3. Here, the measured fragments with Z>2 mainly contribute to the low energy spectra and do not affect the high-energy spectra very much. The predicted ImQMD coalescence-invariant double-ratios for $\gamma_i=2$ change only slightly at low $E_{C.M.}$. On the other hand, the coalescence-invariant double-ratios for $\gamma_i=0.5$ decrease by nearly a factor of two at low $E_{C.M.}$ and approach the no-sensitivity limit of $DR(n/p) \approx 1.2$ as $E_{C.M.}$ decreases. In both cases, the ImQMD calculations at $E_{C.M.}/A>40$ MeV retain sensitivity to the density dependence. Over the whole energy range for both free and coalescence-invariant DR(n/p), the data seem closer to the $\gamma_i=0.5$ calculation but the uncertainties in the measured values are rather large at $E_{C.M.}>40$ MeV, where the effects of cluster emission and secondary decays turn out to be small, as discussed below. More accurate measurements would be needed to distinguish between the $\gamma_i=0.5$ and $\gamma_i=2$ calculations; such measurements should be feasible with a well-designed and dedicated setup.

The previous theoretical studies of $R_{n/p}$ utilized two BUU models, BUU97 [16] and IBUU04 [17] that make no predictions for complex fragment formation. The density dependencies of the symmetry energies employed in IBUU04 (x=0 and x=-1) and BUU97 (F1 and F3) are shown with the dot-dashed and dotted lines in the right panel of Fig. 1. The symmetry energy density dependence of F1 is very similar to that for x=-1 and the symmetry energy density dependence of F3 is softer than that for x=0. More importantly, the IBUU04 code includes: mean field momentum dependencies consistent with the Lane potential, in-medium nucleon-nucleon cross-sections either coinciding with those in free space or incorporating density-dependent modifications that are not included in BUU97. The published BUU97 calculations were performed over a range of impact parameters of b=0-5 fm [16] while the IBUU04 calculations were carried out at b=2 fm [17], as the present calculations. The solid and dashed lines in Figure 4 represent the latest IBUU04 calculations with parameters (x=0 and x=-1) from ref. [17]. Those lines bracket the isospin diffusion data of ref. [14]. The shaded regions represent predictions from BUU97 calculations performed in ref. [16] for two symmetry energy functions, F1 and F3. Irrespectively of the large uncertainties for the BUU97 calculations, it is apparent that the BUU97 results are well in excess of the no-sensitivity limit of DR(n/p)=1.2. Furthermore, it is apparent that far more sensitivity to the symmetry energy is observed for the BUU97 calculations than for the IBUU04 calculations. We do not know the origins of these differences. At high nucleon energies, the results of the present ImQMD calculations for $\gamma_i=0.5$ are similar to the results of momentum independent BUU97 calculations from ref. [16] for the iso-soft (F3) symmetry energy (upper shaded region in Fig. 4). However, the uncertainties in the BUU97 calculations are too large to allow making definitive conclusions.

Lacking clusters, the BUU results must be compared to coalescence-invariant n-p double ratio, DR(n/p), shown as solid stars in Figure 4. The stiffer density-dependent (F1) BUU97 results overlap the data at $E_{C.M.}/A<40$ MeV while the softer density-dependent (F3) results overlap the data at higher energies. In contrast, the IBUU04 calculations lie far below the data near the no-sensitivity limit. Calculations with IBUU04 and a weaker density dependence than (x=0) might be somewhat larger, but the other differences between the transport quantities used in the calculations, such as the cross sections or the effective masses, might contribute more to the differences between the IBUU04 and BUU97 calculations, than do the differences in the symmetry energies. In both calculations, the results from softer density dependence on symmetry energy (x=0 and F3) increase with $E_{C.M.}$, but results from stiffer density dependence (x=-1 and F1) do not. The emission of T=0 alpha clusters enhances the asymmetry of the free nucleons [22], and it seems likely that modeling light clusters would lead to larger values for DR(n/p) in either BUU approach. In any case, the present ImQMD calculations appear to be capable (but both BUU97 and IBUU04 are incapable) of reproducing the energy dependence of the double ratios from low energies, where clusters dominate, to high energies, E/A>40 MeV, where the cluster yields can be neglected. Until this is understood and more accurate data are obtained, the discrepancy with IBUU04 results raises concerns about the extraction of constraints on the symmetry energy from the isospin diffusion data [28]. Additional studies are needed to resolve these issues.

To better understand how the sensitivity of DR(n/p) to the symmetry energy changes with incident energy, we have extracted the calculated excitation function of the double ratios in Figure 5 for high-energy neutrons and protons at incident energies of 35 to 150 MeV per nucleon. Consistent with the forgoing analyses, we show values for DR(n/p) in Fig. 5 for high energy nucleons emitted at $70 \leq \theta_{C.M.} \leq 110$ with $E_{C.M.}$>40 MeV. At all incident energies, the double ratios are larger for $\gamma_i$=0.5 than for $\gamma_i$=2; the largest difference is found at E/A=50 MeV. The values for DR(n/p) for both $\gamma_i$=0.5 and $\gamma_i$=2 and their difference decrease with increasing incident energy. As the importance of collisions and the mean field momentum dependence increases with incident energy, the incident energy dependence of DR(n/p) could provide a useful test of the description of other transport quantities such as effective masses of neutron and protons and the isospin dependence of the in-medium cross-sections.

In summary, we have performed ImQMD transport equation simulations for the systems $^{124}$Sn+$^{124}$Sn and $^{112}$Sn+$^{112}$Sn. Cluster production modifies the spectral double ratios at $E_{C.M}$<40 MeV. The ImQMD model replicates the difference between spectral ratios obtained for free nucleons and those obtained from a coalescence-invariant approach. It also predicts spectral double ratios comparable to the data. However, both the data and calculations at high center of mass energies are not sufficiently accurate to place significant constraints on the density dependence of the symmetry energy. Significant differences are observed between the ImQMD and two BUU models, which need to be resolved before definitive extractions of the density dependence of the symmetry energy from such calculations can be made.

This work has been supported by the U.S. National Science Foundation under Grants PHY-0555893, PHY-0606007, PHY 0216783 (Joint Institute for Nuclear Astrophysics), the High Performance Computing Center (HPCC) at Michigan State University, the Chinese National Science Foundation of China under Grants 10675172, 10175093, 10235030, and the Major State Basic Research development program under contract No. G20000774.

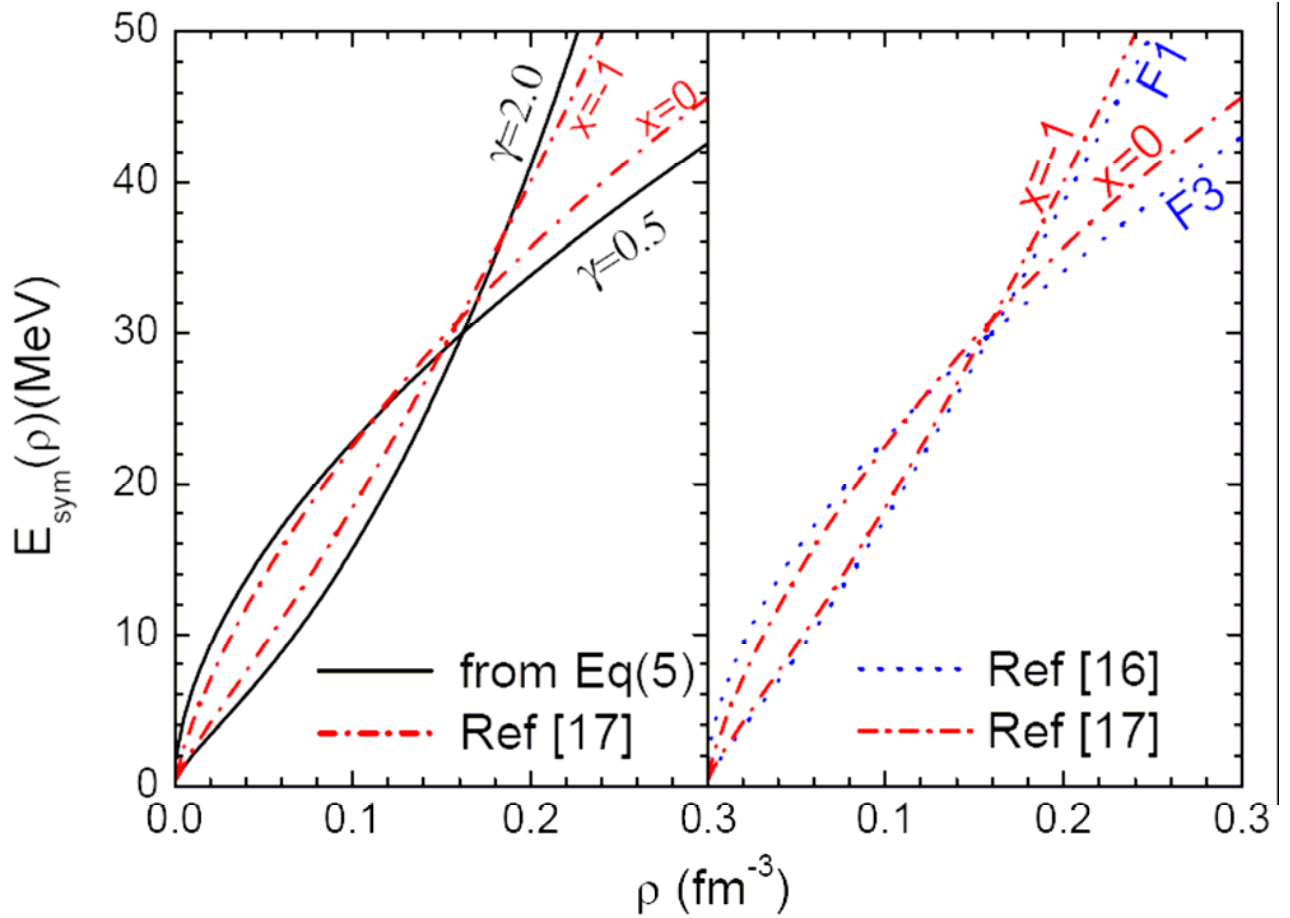

Fig.1: (Color online) Symmetry energy per nucleon plotted as a function of density, The dot-dashed lines labeled x = 0 and x = −1 in both panels represent the symmetry energies used in IBUU04 calculations [17]. The solid lines labeled $\gamma_i=0.5$, and $\gamma_i=2.0$ in the left panel represent the symmetry energies used in the current ImQMD simulations as defined in Equation 5l. The dotted lines in the right panel labeled F1 and F3 represent the symmetry energies used in BUU97 calculations [16].

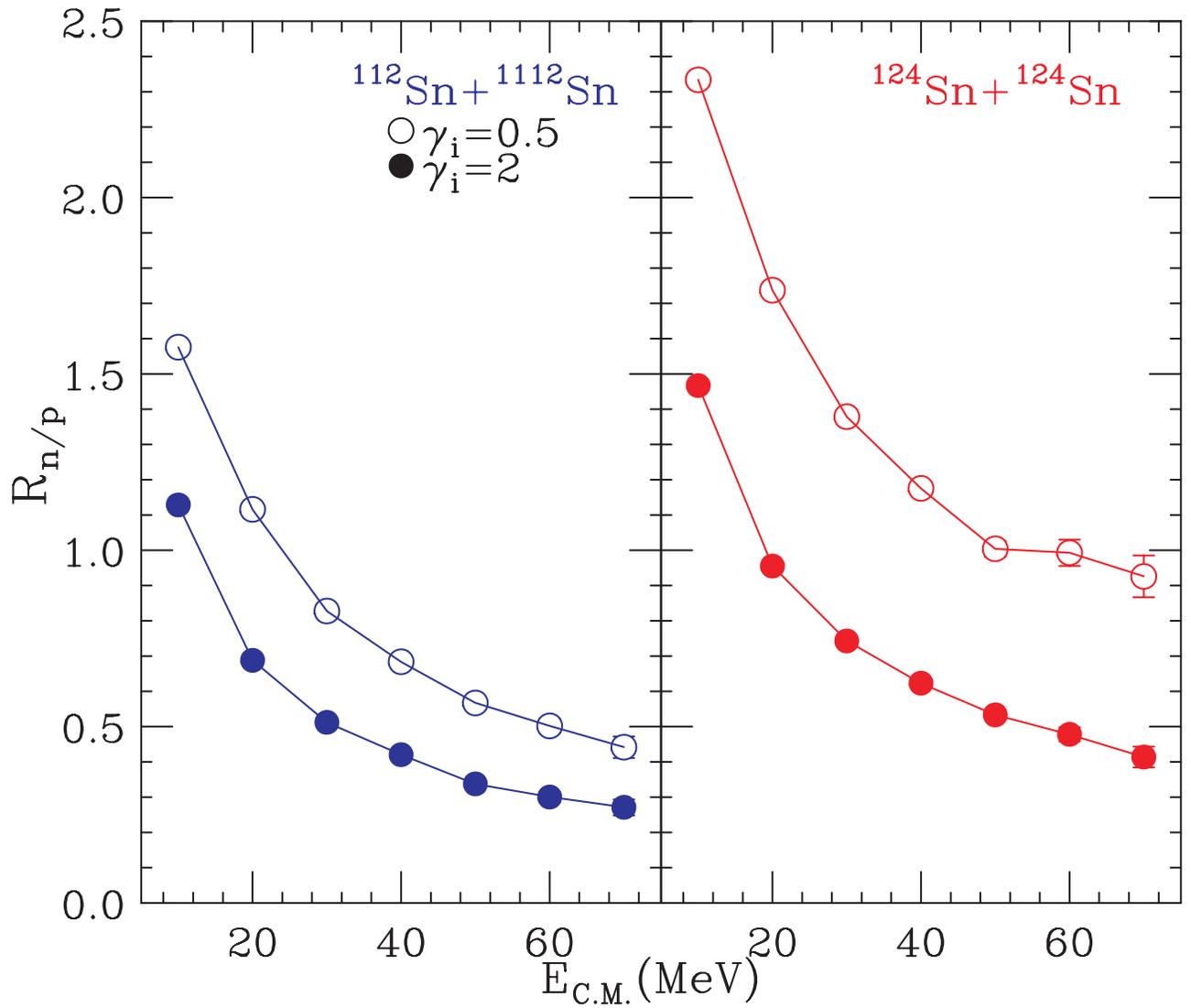

Fig.2: (Color online) The ratio of neutron to proton yields for the $^{112}$Sn+$^{112}$Sn reaction (left panel) and the $^{124}$Sn+$^{124}$Sn reaction (right panel) as a function of the kinetic energy, for free nucleons emitted at $70° \leq \theta_{C.M} \leq 110°$. The open (solid) symbols represent results from simulations using $\gamma_i=0.5$ ($\gamma_i=2.0$) as defined in Equation 5.

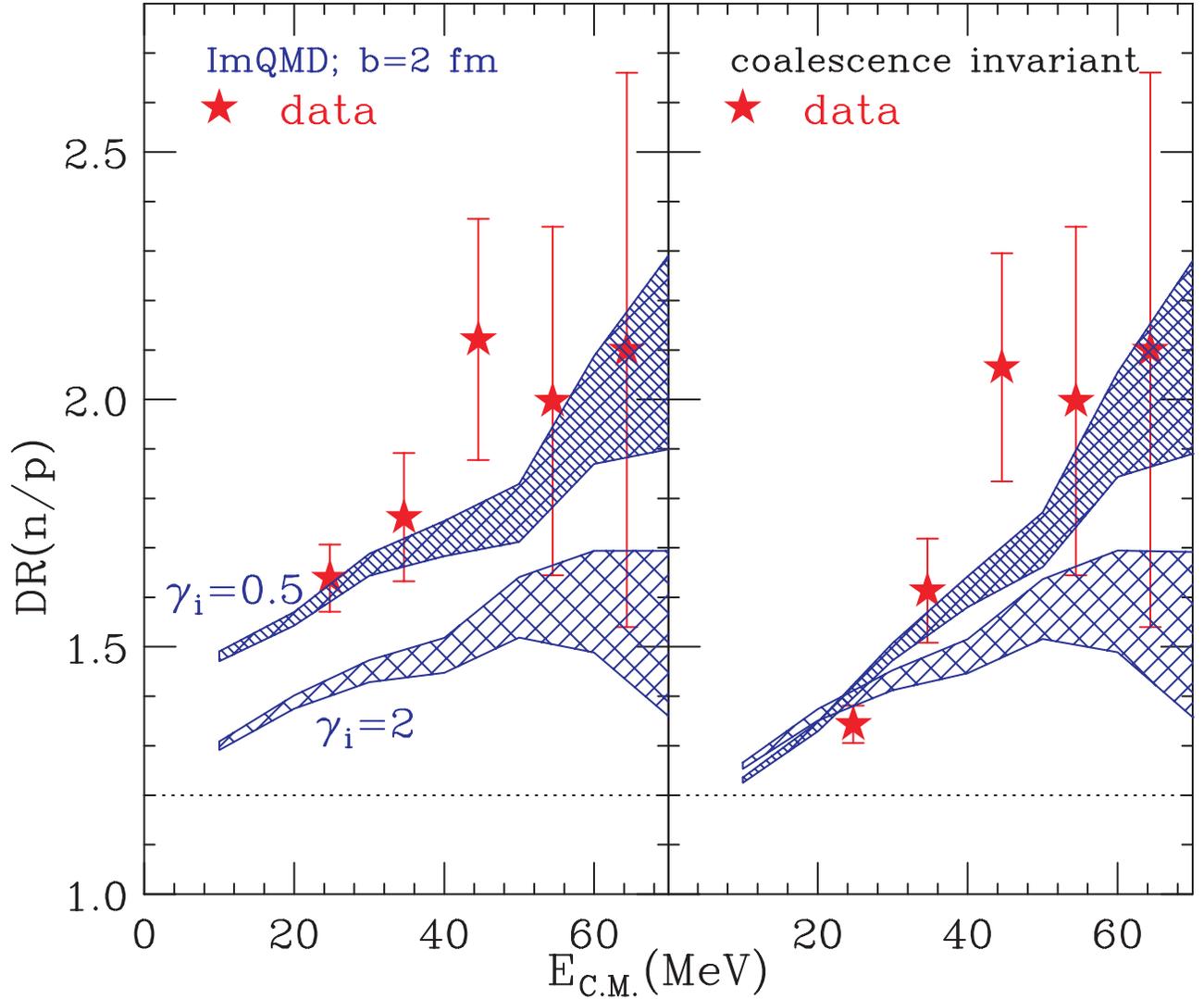

Fig.3: (Color online) The free neutron-proton double-ratio (left panel), and the coalescence-invariant neutron-proton double-ratios (right panel) plotted as a function of kinetic energy of the nucleons. The shaded regions represent calculated results from the ImQMD simulations at b=2 fm. More details are given in the text. The data (solid star points) are taken from Ref [15].

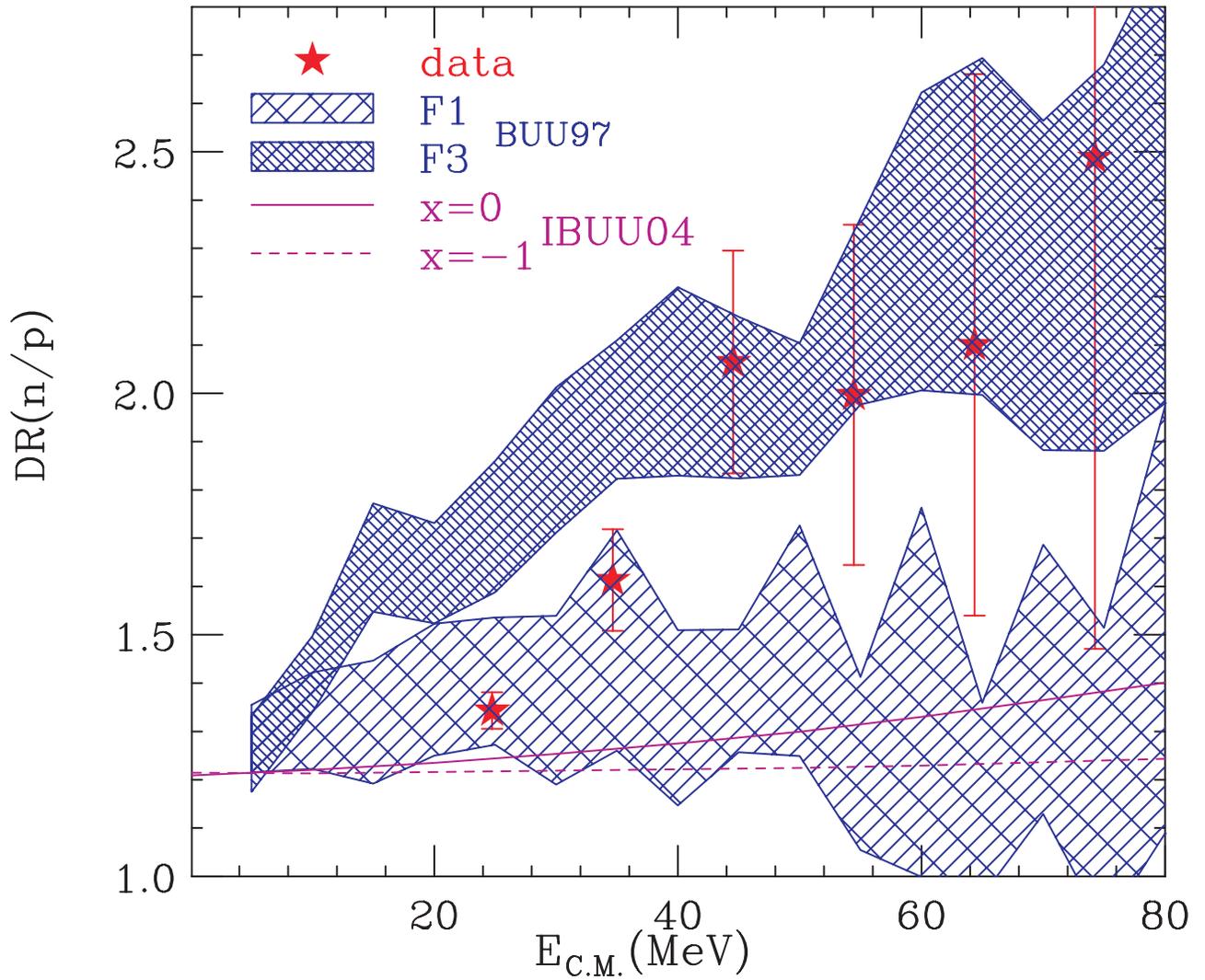

Fig.4: (Color online) Coalescence-invariant neutron-proton double ratios plotted as a function of kinetic energy of the nucleons. The shaded regions represent calculations from the BUU97 simulations taken from ref [16]. The solid and dashed lines represent the results of IBUU04 calculations at b=2 fm, from ref. [17]. The solid stars represent data of Ref [15].

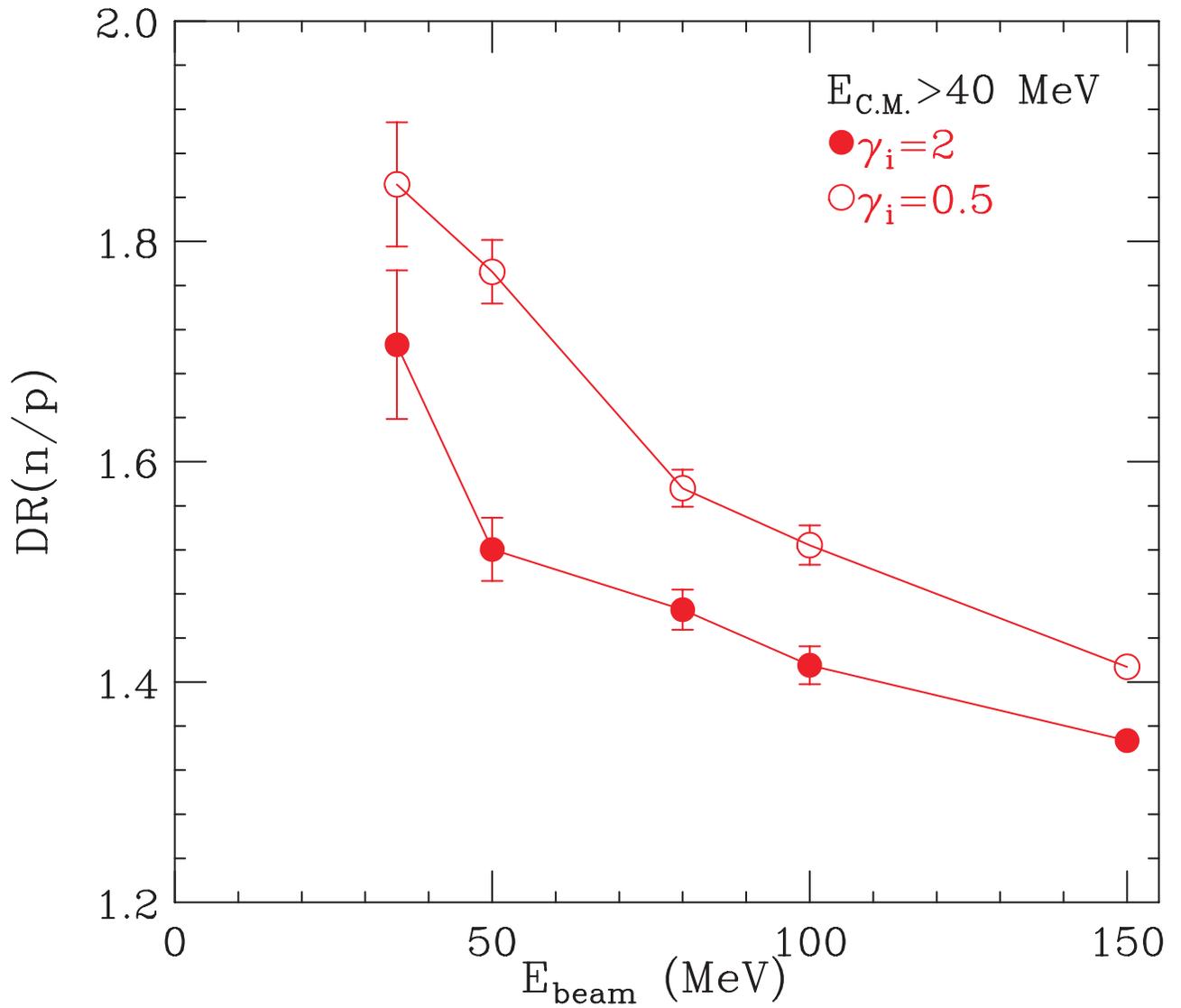

Fig.5: (Color online) Excitation function for neutron- proton double ratios, constructed from high energy ($E_{C.M.}>40$MeV) neutrons and protons, for $\gamma_i=0.5$ (open symbols) and $\gamma_i=2.0$ (solid symbols).